\documentclass[aps,prd,floatfix,preprintnumbers,superscriptaddress,notitlepage,fleqn]{revtex4-1}
\usepackage{dcolumn}
\usepackage{bm}
\usepackage{color}
\usepackage{amssymb}
\usepackage{amsmath}
\usepackage{graphicx}
\usepackage{amsfonts}
\usepackage{slashed}
\usepackage{pstricks}
\usepackage{float}
\usepackage[colorlinks = true,
            linkcolor = blue,
            urlcolor  = blue,
            citecolor = blue,
            anchorcolor = blue]{hyperref}
\usepackage{array}
\allowdisplaybreaks
\usepackage{dcolumn}
\usepackage{epsf}
\usepackage{slashed}
\usepackage{tikz}
\usepackage[compat=1.0.0]{tikz-feynman}
\usepackage{feynmf}
\graphicspath{{figures/}}

\usetikzlibrary{shapes, arrows.meta, chains, scopes, positioning, matrix, calc, external}
\usepackage{color}
\definecolor{gray}{rgb}{0.6,0.6,0.6}
\definecolor{darkgreen}{rgb}{0.0, 0.545098, 0.0}
\definecolor{darkblue}{rgb}{0.0, 0.0, 0.545098}

\mathchardef\mhyphen="2D % hyphen in math mode
\begin{document}
\begin{fmffile}{graphs}
\title{A novel event generator for the automated simulation
of neutrino scattering}
\author{{Joshua} Isaacson}
\author{{Stefan} H\"oche}
\affiliation{
Theoretical Physics Department, Fermi National Accelerator Laboratory, P.O. Box 500, Batavia, IL 60510, USA
}
\author{{Diego} Lopez Gutierrez}
\affiliation{
Physics Department, Harvard University, 17 Oxford Street, Cambridge, MA 02138, USA
}
\author{{Noemi} Rocco}
\affiliation{
Theoretical Physics Department, Fermi National Accelerator Laboratory, P.O. Box 500, Batavia, IL 60510, USA
}
\preprint{FERMILAB-PUB-21-537-T, MCNET-21-31}

%%%%%%%%%%%%%%%%%%%%%%%%%%%%%%%%%%%%%%%%%%%%%%%%%%%%%%%%%%%%%%%%%%%%%%%%%
\begin{abstract} 
An event generation framework is presented that enables the automatic simulation
of events for next-generation neutrino experiments in the Standard Model or extensions thereof. 
The new generator combines the calculation of the leptonic current based on 
an automated matrix element generator, and the computation of the hadronic current 
based on a state-of-the-art nuclear physics model. The approach is validated 
in Standard-Model simulations for electron scattering and neutrino scattering. 
Furthermore, the first fully-differential neutrino trident production results 
are shown in the quasielastic region. 
\end{abstract}
\maketitle
%%%%%%%%%%%%%%%%%%%%%%%%%%%%%%%%%%%%%%%%%%%%%%%%%%%%%%%%%%%%%%%%%%%%%%%%%
\section{Introduction}\label{sec:introduction}

Neutrino physics is entering an era of precision measurements. The development of the 
Deep Underground Neutrino Experiment (DUNE)~\cite{DUNE:2020fgq,DUNE:2020ypp} 
and of the Hyperkamiokande (HyperK)~\cite{Hyper-Kamiokande:2018ofw} detector will allow us 
to probe neutrino interactions to unprecedented precision, and will allow for tests of 
beyond the Standard Model (BSM) scenarios such as proton decay, sterile neutrinos and heavy neutral leptons, 
as well as rare Standard Model (SM) processes such as neutrino tridents~\cite{DUNE:2020fgq, Yano:2021gnf}. 
The excess of electron-like events in measurements at LSND~\cite{PhysRevD.64.112007} 
and MiniBooNE~\cite{PhysRevLett.98.231801,PhysRevLett.102.101802,
  PhysRevLett.105.181801,PhysRevLett.110.161801,MiniBooNE:2018esg,MiniBooNE:2020pnu}
has demonstrated the need for a toolkit which allows the physics community to quickly test 
different hypotheses against experimental data. While calculations can be done on a case-by-case basis, 
such as in Refs.~\cite{Gninenko:2009ks,McKeen:2010rx,Gninenko:2010pr,Dib:2011jh,
  Gninenko:2012rw,Masip:2012ke,Ballett:2016opr,Magill:2018jla,Fischer:2019fbw,Bertuzzo:2018ftf,
  Bertuzzo:2018itn,Ballett:2018ynz,Arguelles:2018mtc,Ballett:2019cqp,Ballett:2019pyw,
  Abdullahi:2020nyr,Abdallah:2020vgg,Abdallah:2020biq,deGouvea:2019qre,Dentler:2019dhz,Chang:2021myh},
a dedicated analysis for each new physics model becomes impractical when there is a wealth of ideas
and a large number of measurements that present simultaneous constraints, such as the recent 
MicroBooNE results~\cite{1953251}. In such cases an end-to-end
simulation should be attempted, starting with the Lagrangian of the underlying new physics hypothesis,
and leading to particle-level events that are generated fully differentially in the many-particle 
phase space by means of Monte-Carlo methods. In addition, it must be possible to implement 
this simulation pipeline as part of experimental analysis frameworks, such that detector effects 
can be fully included.

The accurate simulation of particle-level events based on an underlying physics model has been a cornerstone 
of high-energy physics experiments for many decades. In the context of Large Hadron Collider (LHC) physics, 
the problem of an end-to-end simulation pipeline has been encountered long ago, and it was successfully addressed
through the development of FeynRules~\cite{Christensen:2008py, Alloul:2013bka}, specific event generator
interfaces~\cite{Christensen:2009jx,Christensen:2010wz} and eventually the UFO file format~\cite{Degrande:2011ua}.
A complete toolchain has been developed, starting with the conversion of the Lagrangian to Feynman 
rules~\cite{Christensen:2008py, Alloul:2013bka}, followed by the perturbative calculation of cross sections 
and event simulation by Amegic~\cite{Krauss:2001iv}, Comix~\cite{Gleisberg:2008fv}, Herwig++~\cite{Baehr2008,Bellm:2015jjp}, 
MadGraph~\cite{MadGraph:2014} or Whizard~\cite{Kilian:2007gr}, and finally leading to particle-level event simulation 
with Herwig~\cite{Baehr2008,Bellm:2015jjp}, Pythia~\cite{Sjostrand:2006za,Sjostrand:2014zea} or 
Sherpa~\cite{Gleisberg:2008ta,Sherpa:2019gpd}. The flexibility of the aforementioned tools plays a major role 
in the exploration of model and parameter space by the LHC experiments. In anticipation of similar needs 
at next generation neutrino experiments, we develop a simulation framework that allows for the generation 
of particle-level events in arbitrary new physics models, while at the same time appropriately including 
nuclear effects. We validate our new framework in Standard Model electron- and neutrino-nucleus scattering. 
We also calculate the first results for fully differential neutrino trident production in the quasielastic region.

This manuscript is structured as follows: Section~\ref{sec:TheoryOverview} presents an introduction to the problem 
of neutrino-nucleus scattering and lays out the general framework for the calculation. Section~\ref{sec:COMIX} 
introduces our technique to compute the leptonic tensor, and Sec.~\ref{sec:phasespace} reviews the techniques 
for phase-space integration. Section~\ref{sec:Nuclear} outlines the computation of the hadronic tensor. 
First numerical results are presented in Sec.~\ref{sec:results} and Sec.~\ref{sec:conculsions} contains an outlook.

\section{Theory Overview}\label{sec:TheoryOverview}
In neutrino experiments, the majority of events of interest can be described via an exchange of a single gauge boson. In this case, the differential cross section factorizes as:
\begin{equation}\label{eq:factorization}
    \frac{{\rm d}\sigma}{{\rm d}\Omega} \propto L_{\mu\nu}W^{\mu\nu},
\end{equation}
where $L_{\mu\nu} \, (W^{\mu\nu})$ is the leptonic (hadronic) tensor.

The leptonic tensor can be calculated using methods developed for collider event generation and will be discussed in detail in Sec.~\ref{sec:COMIX}. On the other hand, when dropping spin dependent terms, the hadronic tensor can be written using
the most general Lorentz structure as:
\begin{equation}\label{eq:hadronic_tensor}
    W^{\mu\nu} = \left(-g^{\mu\nu} + \frac{q^\mu q^\nu}{q^2}\right) W_1 + \frac{\hat{p}_a^\mu\hat{p}_a^\nu}{p_a\cdot q} W_2-i\epsilon^{\mu\nu\alpha\beta}\frac{q_\alpha p_{a\beta}}{2p_a\cdot q} W_3,
\end{equation}
where $p_a$ is the momentum of the initial nucleon, $q$ is the momentum of the probe, $\hat{p}_a^\mu = p_a^\mu - \frac{p_a\cdot q}{q^2} q^\mu$, $Q^2 = -q^2$, and $W_i$ are the nuclear structure functions. 

Factorizing the differential cross section in this manner allows us to separate the BSM effects that would be dominant in the leptonic tensor from the complicated nuclear physics calculations of the hadronic tensor. Moreover, neutrino event generators have already implemented many nuclear models into their codes~\cite{Juszczak:2005zs,Andreopoulos:2009zz,Hayato:2009zz,Buss:2011mx}, making the calculation of $W^{\mu\nu}$ a straightforward process. Therefore, we can leverage the work done on the nuclear side to make accurate predictions for different BSM scenarios at DUNE, HyperK, MicroBooNE, and MiniBooNE.
 
While Eq.~\eqref{eq:factorization} works when a single type of gauge boson exchange dominates the total cross section (i.e. the photon for electron scattering, the $W$ boson for charged-current neutrino scattering, and the $Z$ boson for neutral-current neutrino scattering), this approximation breaks down when interference is important. In this case, the differential cross section can be obtained through an extended factorization formula that reads
\begin{equation}\label{eq:factorization2}
    \frac{{\rm d}\sigma}{{\rm d}\Omega} = \sum_{i,j} L^{(ij)}_{\mu\nu}W^{(ij){\mu\nu}},
\end{equation}
where the sum over $i, j$ is for each allowed boson in the process. For example, adding in the $Z$ boson contributions to electron scattering would result in
\begin{equation}\label{eq:photonZ}
    \frac{{\rm d}\sigma}{{\rm d}\Omega} = L^{(\gamma\gamma)}_{\mu\nu}W^{(\gamma\gamma){\mu\nu}} + L^{(\gamma Z)}_{\mu\nu}W^{(\gamma Z){\mu\nu}} +L^{(Z \gamma)}_{\mu\nu}W^{(Z \gamma){\mu\nu}} +L^{(ZZ)}_{\mu\nu}W^{(ZZ){\mu\nu}}.
\end{equation}
This formulation of the problem quickly becomes unwieldy as the number of allowed bosons in the process increases. In these situations, it is better to write the differential cross section as the square of a product of leptonic and hadronic currents:
\begin{equation}\label{eq:factorization3}
    \frac{{\rm d}\sigma}{{\rm d}\Omega} = \bigg|\sum_{i} L^{(i)}_{\mu} W^{(i)\mu}\bigg|^2,
\end{equation}
where the interference terms are automatically included by taking the square of the sum of amplitudes. In this work, we will use Eq.~\eqref{eq:factorization3} for our calculations, but in the cases when the nuclear physics cannot be written as a current, we provide the option to work with the leptonic tensor in Eq.~\eqref{eq:factorization2}.

The BSM calculations we intend to automate involve only the leptonic current and are independent of the exact nuclear model. 
We will therefore focus on the quasielastic region in this work for concreteness. 
Specifically, the hadronic current will be computed in the impulse approximation (IA),
using the spectral function formalism as discussed in Refs.~\cite{Benhar:2006wy, Rocco:2018mwt}. 
In this approximation, the nuclear current can be expressed as a sum of one-body currents, and the nuclear final state 
can be written as a struck nucleon and an $A-1$ spectator nuclear remnant. Given the initial nuclear ground state 
$\lvert\psi_0^A\rangle$ and the nuclear final state $\langle\psi_f^A\rvert$, the current operator can
be rewritten through the insertion of a complete set of states ($\sum_k \lvert k \rangle \langle k \rvert$) in the IA as:
\begin{equation}
    W^\mu = \langle \psi_f^A\rvert \mathcal{J}^{\mu} \lvert \psi_0^A\rangle \rightarrow \sum_k \left[\langle \psi_f^{A-1}\rvert \otimes \langle k \rvert\right]\rvert \psi_0^A\rangle \langle p_a + q \rvert \sum_i \mathcal{J}_i^{\mu} \lvert k \rangle,
\end{equation}
where $\mathcal{J}^{\mu}$ is the current operator of the boson probing the nucleus, and $\mathcal{J}_i^{\mu}$ is the current operator contribution from the $i^{\rm th}$ nucleon in the IA.
The incoherent contribution to the hadronic tensor can then be calculated as:
\begin{equation} \label{eq:spectral}
    W^{\mu\nu} = \int \frac{{\rm d}^3 p_a}{(2\pi)^3} {\rm d} E\, S(\mathbf{p_a}, E_r) \sum_i \langle p_a \rvert \mathcal{J}_i^{\mu}\lvert p_a + q\rangle \langle p_a + q \rvert \mathcal{J}_i^{\nu\dagger}\lvert k \rangle \delta\left(\omega -E_r +m - E'\right),
\end{equation}
where $\omega$ is the lepton energy loss, $E_r$ is the excitation energy of the $A-1$ nucleon system, $E'$ is the outgoing nucleon energy and $S(\mathbf{p}_a, E_r)$ is the
spectral function which gives the probability of removing a nucleon with three-momentum $\mathbf{p}_a$ and
excitation energy $E_r$. The spectral functions of finite nuclei have been obtained using different nuclear many-body approaches~\cite{Dickhoff:2018wdd,Dickhoff:2019zmo,Barbieri:2009nx,Benhar:1994hw,Sick:1994vj}. In particular, those computed within the correlated-basis function (CBF) theory, Self Consistent Green's Function (SCGF), and variational Monte Carlo (VMC) methods have yielded consistent inclusive lepton-scattering cross sections on different nuclear targets~\cite{Rocco:2018mwt,Rocco:2018vbf,Andreoli:2021cxo,Ankowski:2014yfa}. 
The CBF spectral function used in this work is given as a sum of two terms. The first term, which describes the low excitation energy and momentum region, uses as input spectroscopic factors extracted from ($e,e'p$) scattering measurements. The second term accounts for unbound states of the $A-1$ spectator system in which at least one of the spectator nucleons is in a continuum state. This contribution is obtained by folding the correlation component of the nuclear matter spectral function obtained within the CBF theory with the the density profile of the nucleus~\cite{Benhar:1989ue}. 

The nuclear one-body current operator can be expressed in general as the sum of the vector (V) and axial-vector (A) terms given as:
\begin{align} \label{eq:currents}
    \mathcal{J}^\mu &= \left(\mathcal{J}_{V}^{\mu} + \mathcal{J}_{A}^{\mu}\right) \nonumber \\
    \mathcal{J}^\mu_{V} &= \gamma^\mu \mathcal{F}^{a}_1 + i\sigma^{\mu\nu} q_{\nu} \frac{\mathcal{F}^{a}_2}{2M} \\
    \mathcal{J}^\mu_{A} &= -\gamma^\mu\gamma_5 \mathcal{F}^{a}_A - q^{\mu}\gamma_5\frac{\mathcal{F}^{a}_P}{M} \nonumber,
\end{align}
where $\mathcal{F}^{a}_i$ are the process dependent nuclear form factors for the exchange of the boson $a$ and $M$ is the nucleon mass. Since the pseudo-scalar form factor ($\mathcal{F}^{a}_P$) is multiplied by the mass of the outgoing lepton in the cross section, we neglect the
contribution from it in this work. The values for the form factors considered here are detailed in Sec.~\ref{sec:Nuclear}.

\section{Calculation of the Leptonic Current}\label{sec:COMIX}
Employing a dedicated version of the general-purpose event generator
Sherpa~\cite{Gleisberg:2003xi,Gleisberg:2008ta,Sherpa:2019gpd},
we construct an interface to the Comix matrix element
generator~\cite{Gleisberg:2008fv} to extract the leptonic current.
In Comix, the matrix element is computed using the color-dressed Berends-Giele 
recursive relations~\cite{Duhr:2006iq}, which can be understood as a 
Dyson-Schwinger based technique~\cite{Kanaki:2000ey,Mangano:2002ea}.
In this technique, the full tree-level scattering amplitude is determined from off-shell currents 
which are composed of all sub-diagrams connecting a certain sub-set 
of external particles. These currents depend 
on the momenta $p_1,\ldots,p_n$ of external particles $1,\ldots, n$, 
and on their helicities and colors.

The off-shell currents satisfy the recursive relations
\begin{equation}\label{eq:sm_general_recursion}
  J_{\alpha}(\pi)=P_{\alpha}(\pi)\,
    \sum\limits_{\mathcal{V}_{\alpha}^{\;\alpha_1,\,\alpha_2}}
    \sum\limits_{\mathcal{P}_2(\pi)}\mathcal{S}(\pi_1,\pi_2)\;
    \mathcal{V}_{\alpha}^{\,\alpha_1,\,\alpha_2}(\pi_1,\pi_2)\,
    J_{\alpha_1}(\pi_1)J_{\alpha_2}(\pi_2)\;,
\end{equation}
where, $P_{\alpha}(\pi)$ denotes a propagator, which 
depends on the particle type $\alpha$ and the set of particles, $\pi$.
The term $\mathcal{V}_{\alpha}^{\,\alpha_1,\alpha_2}(\pi_1,\pi_2)$ 
represents a vertex, which depends on the particle types $\alpha$, 
$\alpha_1$ and $\alpha_2$ and the decomposition of the set of 
particle labels, $\pi$, into disjoint subsets $\pi_1$ and $\pi_2$.
The quantity $\mathcal{S}(\pi_1,\pi_2)$ is the symmetry factor 
associated with the decomposition of $\pi$ into $\pi_1$ and $\pi_2$~\cite{Gleisberg:2008fv}.
Superscripts refer to incoming 
particles at the vertex, subscripts to outgoing particles.
The sums run over all vertices in the interaction model and
over all unordered partitions $\mathcal{P}_2$ of the set $\pi$
into two disjoint subsets.
Since we will use the above recursion only to compute the leptonic current, 
we can suppress color indices. Helicity labels are implied.
The initial currents $J_{\alpha_i}(p_i)$ can be
determined based on the spin of the $i^{\text{th}}$ external particle. The
external currents are given by:
\begin{equation}
    J_\alpha(p_i) = 
    \begin{cases}
        1 & \text{spin} = 0 \\
        u(p_i)\ \text{or}\ v(p_i) & \text{spin} = 1/2 \\
        \epsilon_\alpha(p_i, k) & \text{spin} = 1
    \end{cases},
\end{equation}
where $u(p_i)$ and $v(p_i)$ are the solutions to the Dirac equation, 
$\epsilon_\alpha$ is the polarization
vector, and $k$ is an auxiliary vector to define the polarization.

A complete list of interaction vertices 
for the Standard Model can be found in Ref.~\cite{Gleisberg:2008fv}.
For completeness, we also list the nontrivial expressions for the natively 
implemented Lorentz structures in App.~\ref{sec:lorentz_structures}.
Comix allows to include Feynman rules for nearly arbitrary new physics
scenarios into Eq.~\eqref{eq:sm_general_recursion} by means of
a generic interface to the UFO file format~\cite{Degrande:2011ua}, 
and an automated generator for the Lorentz (and color) 
structures~\cite{Hoche:2014kca,Krauss:2016ely}. This extension of Comix to BSM calculations
also discusses the extension of Eq.~\eqref{eq:sm_general_recursion} from containing only 3-point
vertices to arbitrary $N$-point vertices.
The generation of UFO files
can be accomplished through the use of the FeynRules program~\cite{Christensen:2008py,Alloul:2013bka}, which
can take any Lagrangian and obtain the needed Feynman rules for tree-level
calculations.

In order to compute the leptonic current needed to evaluate 
Eq.~\eqref{eq:factorization}, we introduce point-like nucleons
into the theory, which act as auxiliary particles that are needed 
only for bookkeeping purposes and to construct a formally complete
scattering amplitude.
The off-shell currents coupling  to these nucleons are then used
to define the leptonic current or the leptonic tensor as
\begin{equation}
\begin{split}
    L_{\mu}^{(i)}(1,\ldots,m)=&\;J^{(i)}_\mu(1,\ldots,m)\;,\\
    L_{\mu\nu}^{(ij)}(1,\ldots,m)=&\;J^{(i)}_\mu(1,\ldots,m)J^{(j)\dagger}_\nu(1,\ldots,m)\;,
\end{split}
\end{equation}
where the particles $1,\ldots,m$ are the non strongly interacting 
particles that contribute to the leptonic current.

Through the use of the recursion relations, and the interface with UFO files generated from FeynRules,
this generator can calculate almost any leptonic current that may be of interest. The only limitations
on the leptonic currents that can be calculated are: can not currently handle any spin $>$ 1 particles,
can not have any color charged particles, only spin-1 particles can probe the nucleus, and only tree-level diagrams can be calculated. Of these limitations,
only the exclusion of the color charged particles can not be resolved with future work. Allowing color charged
particles to interact with the nucleus breaks the assumption that the degrees of freedom in QCD are protons
and neutrons. Implementing other probes of the nucleus involves updating the nuclear physics to include the
appropriate form factors for the different spin probes. Handling particles with spin $>$ 1 requires implementing
the needed external particle states and appropriate propagators. The automation of one-loop diagrams has been discussed in detail in Refs.~\cite{Hirschi:2011pa,Denner:2017wsf,Buccioni:2019sur}.

\section{Phase-space integration}\label{sec:phasespace}
We employ the recursive phase-space generation techniques 
of Ref.~\cite{Byckling:1969sx} in combination with the multi-channel
method of Ref.~\cite{Kleiss:1994qy} and the adaptive multi-dimensional
integration algorithm Vegas~\cite{Lepage:1980dq} in order to perform
the phase-space integrals.
Considering a $2\to n$ scattering process with incoming particles
$a$ and $b$ and outgoing particles $1\ldots n$, the $n$-particle
differential phase space element reads
\begin{equation}\label{eq:phasespace}
  {\rm d}\Phi_n(a,b;1,\ldots,n)=
    \left[\,\prod\limits_{i=1}^n\frac{{\rm d}^4 p_i}{(2\pi)^3}\,
    \delta(p_i^2-m_i^2)\Theta(p_{i0})\,\right]\,
    (2\pi)^4\delta^{(4)}\left(p_a+p_b-\sum_{i=1}^n p_i\right)\;,
\end{equation}
where $p_i$ and $m_i$ are the momentum and on-shell masses of outgoing particles, respectively.

For concreteness, consider the $2 \to 2$ process $l + {}^{12}C \to l' + N + X$ in the quasielastic regime.
In this case, we not only have to consider the final state lepton and nucleon, but also the initial
state of the system. In the quasielastic regime, the initial momentum of the nucleon can be generated
by considering an isotropic three-momentum and a removal energy of the nucleon inside the nucleus.
Furthermore, if the lepton is not monochromatic, then the momentum of the lepton can be generated
according to the initial flux. Putting all the components together, the full differential phase space
element is given as ${\rm d}\Phi_2(a,b;1,2){\rm d}^3 \mathbf{p}_a {\rm d} E_r {\rm d}^3 \mathbf{p}_b$, where ${\rm d}\Phi_2(a,b;1,2)$ 
is given in Eq.~\eqref{eq:phasespace}, $\mathbf{p}_a (E_r)$ is the three-momentum (removal energy) of the initial nucleon,
and $\mathbf{p}_b$ is the three-momentum of the initial lepton. This process will proceed through a $t$-channel 
exchange of a gauge boson. Therefore, it is efficient to rewrite the two-body final state phase space
such that it appropriately samples a $t$-channel momentum distribution as:
\begin{equation}\label{eq:t-channel}
    {\rm d}\Phi_2(a,b;1,2) = \frac{\lambda\left(s_{ab},s_{1}, s_{2}\right)}{16\pi^2 2s_{ab}}\,
    {\rm d}\cos\theta_{1}{\rm d}\phi_{1},
\end{equation}
where $\theta_1 (\phi_1)$ is the polar (azimuthal) angle with respect to the axis formed by $p_a+p_b$.
Here we have introduced the K\"allen function:
\begin{equation}\label{eq:Kallen}
\lambda^2\left(s_a,s_b,s_c\right)={\left(s_a - s_b - s_c\right)^2-4s_bs_c},
\end{equation}
where $s_i$ denotes the invariant mass of the particles $i$. The differential phase-space element
${\rm d}^3\mathbf{p}_a$ can be rewritten as:
\begin{equation}\label{eq:initial_nucleon}
    {\rm d}^3 \mathbf{p}_a = |\mathbf{p}_a|^2 {\rm d}\mathbf{p}_a {\rm d}\cos\theta_a {\rm d}\phi_a,
\end{equation}
where $E_r$ is the removal energy, $\theta_a (\phi_a)$ is the polar (azimuthal) angle with respect to the beam direction, and the energy of the nucleon is given as $E_a=m_{N} - E_r$ to convert the 
quasielastic energy conservation $\delta$-function given in Eq.~\eqref{eq:spectral} to $\delta\left(E_a+E_b-E_1-E_2\right)$, which
is consistent with the energy conservation of Eq.~\eqref{eq:phasespace}. The initial distribution for the lepton depends
on the experiment being considered, and is typically given as a probability distribution given some initial momentum
(and position in the case of fluxes from NuMI beam simulations~\cite{Adamson:2015dkw}).
Events can be generated according to this probability distribution with various MC techniques. 
In this work, we consider only monochromatic beams, so details of these methods are left to a future work.

To generalize the phase-space integration to higher multiplicities,
we follow Ref.~\cite{James:1968gu}, factorizing Eq.~\eqref{eq:phasespace} as
\begin{equation}\label{eq:split_ps}
  {\rm d}\Phi_n(a,b;1,\ldots,n)=
    {\rm d}\Phi_{n-m+1}(a,b;\pi,m+1,\ldots,n)\,\frac{{\rm d} s_\pi}{2\pi}\,
    {\rm d}\Phi_m(\pi;1,\ldots,m)\;.
\end{equation}
In this context, $\pi=\{1,\ldots,m\}$ corresponds to a subset of particle indices,
similar to the notation in Sec.~\ref{sec:COMIX}. We use overlined letters 
to denote the missing subset, e.g.\ $\overline{\alpha}=\{a,b,1,\ldots,n\}\setminus\alpha$.
Equation~\eqref{eq:split_ps} allows the decomposition of the complete phase space
into building blocks corresponding to the propagator-like term ${\rm d} s_\pi/(2\pi)$
and the $s$- and $t$-channel decay processes
\begin{equation}\label{eq:ps_building_blocks}
  \begin{split}
    S_{\pi}^{\,\rho,\pi\setminus\rho}&=
     \frac{\lambda(s_\pi,s_\rho,s_{\pi\setminus\rho})}{
     16\pi^2\,2\,s_{\pi}}\;{\rm d}\cos\theta_\rho\,{\rm d}\phi_\rho\;,\\
    T_{\alpha,b}^{\,\pi,\overline{\alpha b\pi}}&=
     \frac{\lambda(s_{\alpha b},s_\pi,
       s_{\;\overline{\alpha b\pi}})}{16\pi^2\,2s_{\alpha b}}\;
     {\rm d}\cos\theta_\pi\,{\rm d}\phi_\pi\;,\\
  \end{split}
\end{equation}
where $T_{\alpha,b}^{\,\pi,\overline{\alpha b\pi}}$
is the same function as in Eq.~\eqref{eq:t-channel}.
In addition, an overall factor of $(2\pi)^4\,{\rm d}^4 p_{\,\overline{\alpha b}}\;
      \delta^{(4)}(p_\alpha+p_b-p_{\,\overline{\alpha b}})$ 
guarantees four-momentum conservation. The basic idea of Ref.~\cite{Byckling:1969sx}
is to match the indices in the virtuality integrals of Eq.~\eqref{eq:split_ps} 
and the lower left indices in the $T$ functions to indices of $t$-channel currents
in the hard matrix element. In this manner, one obtains an optimal integrator
for a particular Feynman diagram in a scalar theory that contains propagators
with all these indices. An optimal integrator for {\it all} scalar diagrams
can then be constructed using the multi-channel technique~\cite{Kleiss:1994qy}.
Finally, the spin dependence of the hard matrix element can be mapped out
more carefully with the help of adaptive MC algorithms~\cite{
  Lepage:1977sw,Bothmann:2020ywa,Gao:2020zvv}.

\section{Nuclear Matrix Element Interface}\label{sec:Nuclear}
While the UFO file is extremely flexible, it is missing a way to easily define the nuclear form factors involved within the interaction. We propose to use an extension to the UFO file format that provides a way to consistently interface with the form factors used in the neutrino event generators and other neutrino-nucleus interaction codes. This extension will only be valid as long as the Conserved Vector Current (CVC) hypothesis is valid. The CVC provides a method to relate the vector form factor for an arbitrary model to the
electromagnetic form factors ($F_1, F_2$). The electomagnetic form factors $F_1$ and $F_2$ can
be defined in terms of the electric and magnetic form factors as:
\begin{align}
    F_1^{p,n} &= \frac{G_E^{p,n} + \tau G_M^{p,n}}{1+\tau} \nonumber \\
    F_2^{p,n} &= \frac{G_E^{p,n} - G_M^{p,n}}{1+\tau},
\end{align}
with $\tau = -q^2/(2M)^2$. In this work, we consider the Kelly parameterization for the electric and magnetic form factors~\cite{Kelly:2004hm}.

Additionally, we will consider a global axial-vector form factor as well. While it is straightforward
to implement any form factor, such as the ``z-expansion"~\cite{Meyer:2016oeg}, we choose to use the dipole parameterization to allow for a fair comparison to Ref.~\cite{Rocco:2018mwt}.
The standard dipole parameterization is given as
\begin{equation}
    F_A = \frac{g_A}{(1-q^2/M_A^2)^2}, 
\end{equation}
where the nucleon axial-vector coupling constant is taken to be $g_A = 1.2694$ and the axial mass
is taken to be $M_A$ = 1.0 GeV. Studying the impact of different
parameterizations of the axial-vector form factor are beyond the scope of this work.

The form factors for an arbitrary process can be expressed in the form factors given above through the use of their isospin dependent interactions. Here we work out the
form factors for the Standard Model photon, $Z$ boson, and $W$ boson, but the process
can be applied similarly to any BSM scenario. Since the form factors are defined in terms of the coupling to the photon, the photon form factors ($\mathcal{F}^{\gamma}_i$; see Eq.~\eqref{eq:currents}) are straightforward, and
given as:
\begin{equation}
    \mathcal{F}^{\gamma (p,n)}_i = F^{p,n}_i, \quad\quad \mathcal{F}^{\gamma}_A = 0.
\end{equation}
The $W^{\pm}$ boson couples via isospin through the $\tau^{\pm}$ operator, which leads to the form factors:
\begin{equation}
    \mathcal{F}^{W (p, n)}_i = F^{p}_i - F^{n}_i, \quad\quad \mathcal{F}^{W}_A = F_A.
\end{equation}
The $Z$ boson also couples via isospin, and the form factors are
\begin{align}
   \mathcal{F}_i^{Z (p)} &= \left(\frac{1}{2} - 2\sin^2\theta_W\right)F_i^{p} - \frac{1}{2} F_i^{n}, \quad\quad\mathcal{F}_A^{Z (p)} = \frac{1}{2} F_A, \nonumber \\
   \mathcal{F}_i^{Z (n)} &= \left(\frac{1}{2} - 2\sin^2\theta_W\right)F_i^{n} - \frac{1}{2} F_i^{p}, \quad\quad\mathcal{F}_A^{Z (n)} = -\frac{1}{2} F_A,
\end{align}
where $\sin\theta_W$ is the weak mixing angle. We leave the examples of working out the form factors for specific BSM scenarios to a future work.

\section{Results}\label{sec:results}

We consider the processes of electron and neutrino scattering off Carbon, and
neutrino trident production off Carbon. The electron scattering and neutrino scattering
are used as a means to validate the code. The trident production is the first fully-differential calculation of tridents
in the quasielastic region using nuclear spectral functions, to demonstrate the ability to extend the
predictions beyond $2 \to 2$ processes. For simplicity, we only consider a monochromatic beam for the
incoming particle, but it is straightforward to include appropriate fluxes.
Additionally, we do not include any final state interaction effects, which are known to shift the location of the peak of the distribution, reduce the height of the peak and broaden the tails of the distribution. For all the calculations, we set
our electroweak parameters in the $(\alpha, G_F, M_Z)$-scheme, and the values are
chosen as $\alpha = 1/137$, $G_F=1.16637\times 10^{-5}$ GeV${}^{-2}$, and $M_Z=91.1876$ GeV. Throughout the calculations, all leptons are considered massless particles. Furthermore, we include the full propagator and do not take the infinite mass limit for the contributions from $W$ or $Z$ bosons. This differs from previous works in which the four-point Fermi interactions are used to model the neutrino-nucleus interactions.

\subsection{Electron Scattering}\label{subsec:electron}

There is a plethora of electron scattering data on a Carbon target that has been collected. Here we 
compare to a select few experimental results, along with the predictions based on Ref.~\cite{Rocco:2019gfb}. The data
is the double differential cross section in solid angle and the energy transfer ($\omega$) versus the energy transfer to the nucleus.

\begin{figure}[!ht]
    \centering
    \includegraphics[width=0.7\textwidth]{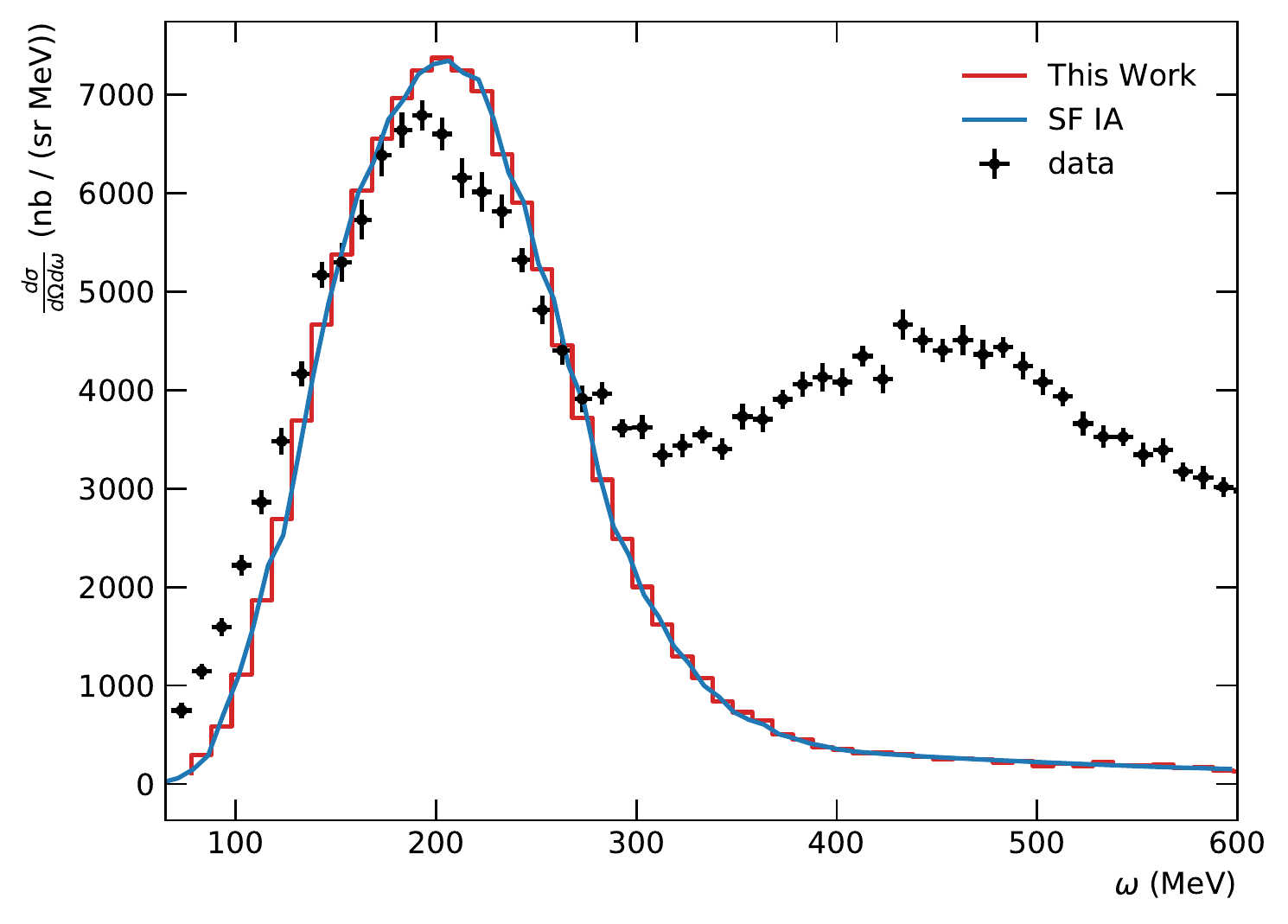}
    \caption{A comparison between our calculation, the data from Ref.~\cite{Sealock:1989nx}, and the 
    theory calculation based on Ref.~\cite{Rocco:2019gfb} for a 961 MeV electron scattering off Carbon at an angle of 37.5 degrees. The two theory calculations are consistent with each other, and the difference with the data can be explained by including final state interaction effects.}
    \label{fig:eC_961_37p5}
\end{figure}

First, we consider the electron scattering off Carbon with an electron beam energy of 961 MeV
and 37.5 degrees scattering angle~\cite{Sealock:1989nx}. The comparison between the data, the results based on 
Ref.~\cite{Rocco:2019gfb}, and our work can be seen in Fig.~\ref{fig:eC_961_37p5}. We observe that the calculation is consistent with the original theory calculation. Furthermore, the difference in the quasielastic peak region between the theory calculation and the data can be explained by including final state interaction effects, as shown in Ref.~\cite{Rocco:2019gfb}. The high energy transfer tail can be explained through the inclusion of two-body currents, resonance production, and deep-inelastic scattering~\cite{Rocco:2019gfb}.

\begin{figure}
    \centering
    \includegraphics[width=0.7\textwidth]{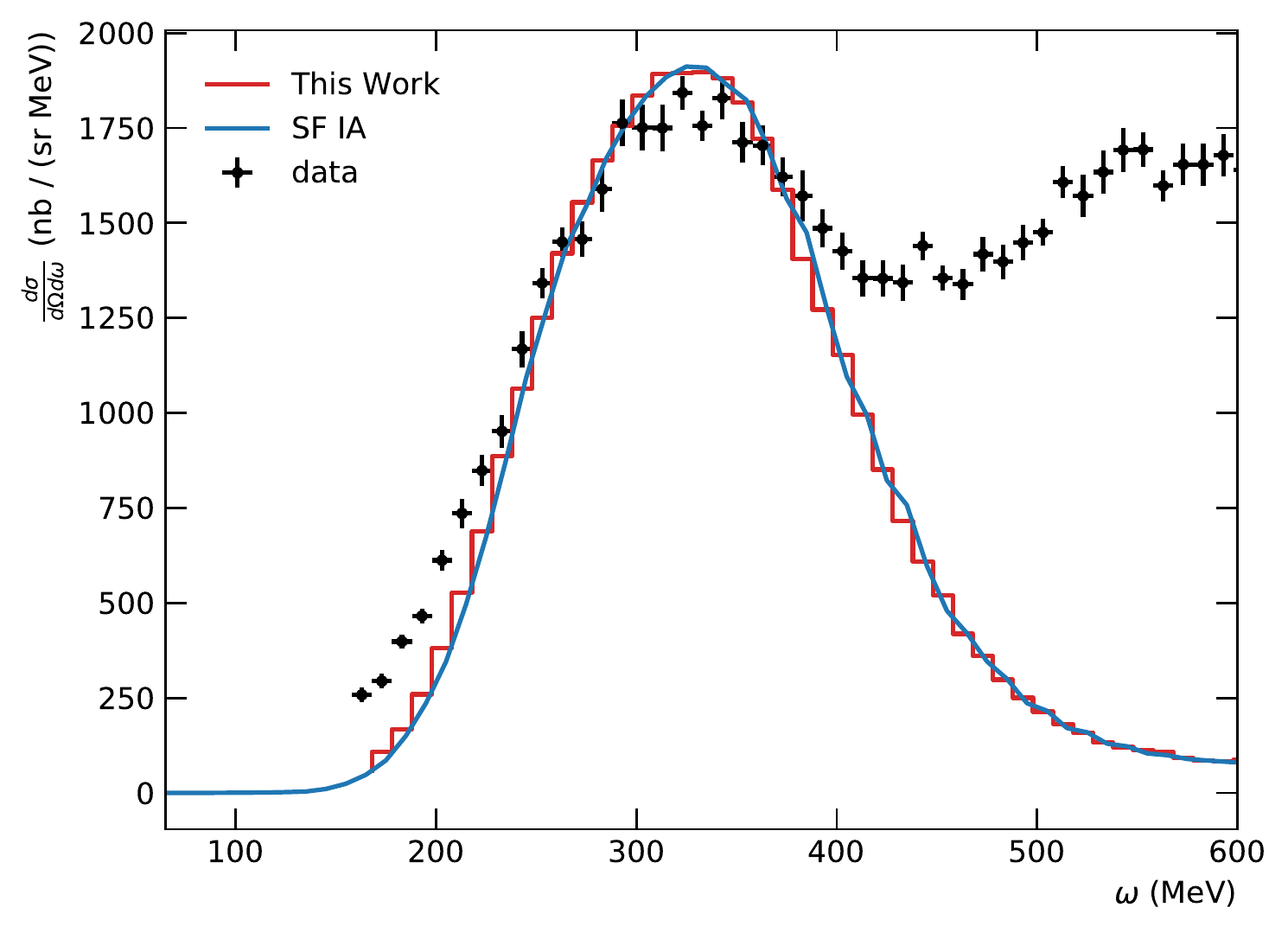}
    \caption{A comparison between our calculation, the data from Ref.~\cite{Sealock:1989nx}, and the 
    theory calculation based on Ref.~\cite{Rocco:2019gfb} for a 1300 MeV electron scattering off Carbon at an angle of 37.5 degrees. The two theory calculations are consistent with each other, and the difference with the data can be explained by including final state interaction effects.}
    \label{fig:eC_1300_37}
\end{figure}

Figure~\ref{fig:eC_1300_37} displays the comparison between the two theory approaches and data for electron scattering off Carbon with an electron beam energy of 1300 MeV and 37.5 degrees scattering angle~\cite{Sealock:1989nx}. Again, we observe that the two theory calculations are consistent. There is an overall agreement between the theory prediction based on Ref.~\cite{Rocco:2019gfb} and the data in the quasielastic region, the small difference at low $\omega$ has to be ascribed to final state interaction effects as in Fig.~\ref{fig:eC_1300_37}. These can be accounted for as demonstrated in Ref.~\cite{Rocco:2019gfb} but their inclusion is beyond the scope of this work. 

\subsection{Neutrino Scattering}\label{subsec:neutrino}

For neutrino scattering, we consider both neutrino and anti-neutrino beams. Since there are no high-energy monochromatic beams of neutrinos, we only compare our results to those
from Ref.~\cite{Rocco:2018mwt}, but do not compare to any experimental data. For both neutrino and anti-neutrino
beams, we consider both total cross section as a function of incoming neutrino energy, along with the
double differential cross section in outgoing angle and energy of the outgoing lepton versus the energy
transfer to the nucleus. We consider only charged-current (CC) interactions here. In order to compare to the results of Ref.~\cite{Rocco:2018mwt}, we include the effects from Pauli blocking by ensuring that the outgoing nucleon has a momentum greater than the Fermi momentum $k_F= 225$ MeV.

\begin{figure}[!ht]
    \centering
    \includegraphics[width=0.45\textwidth]{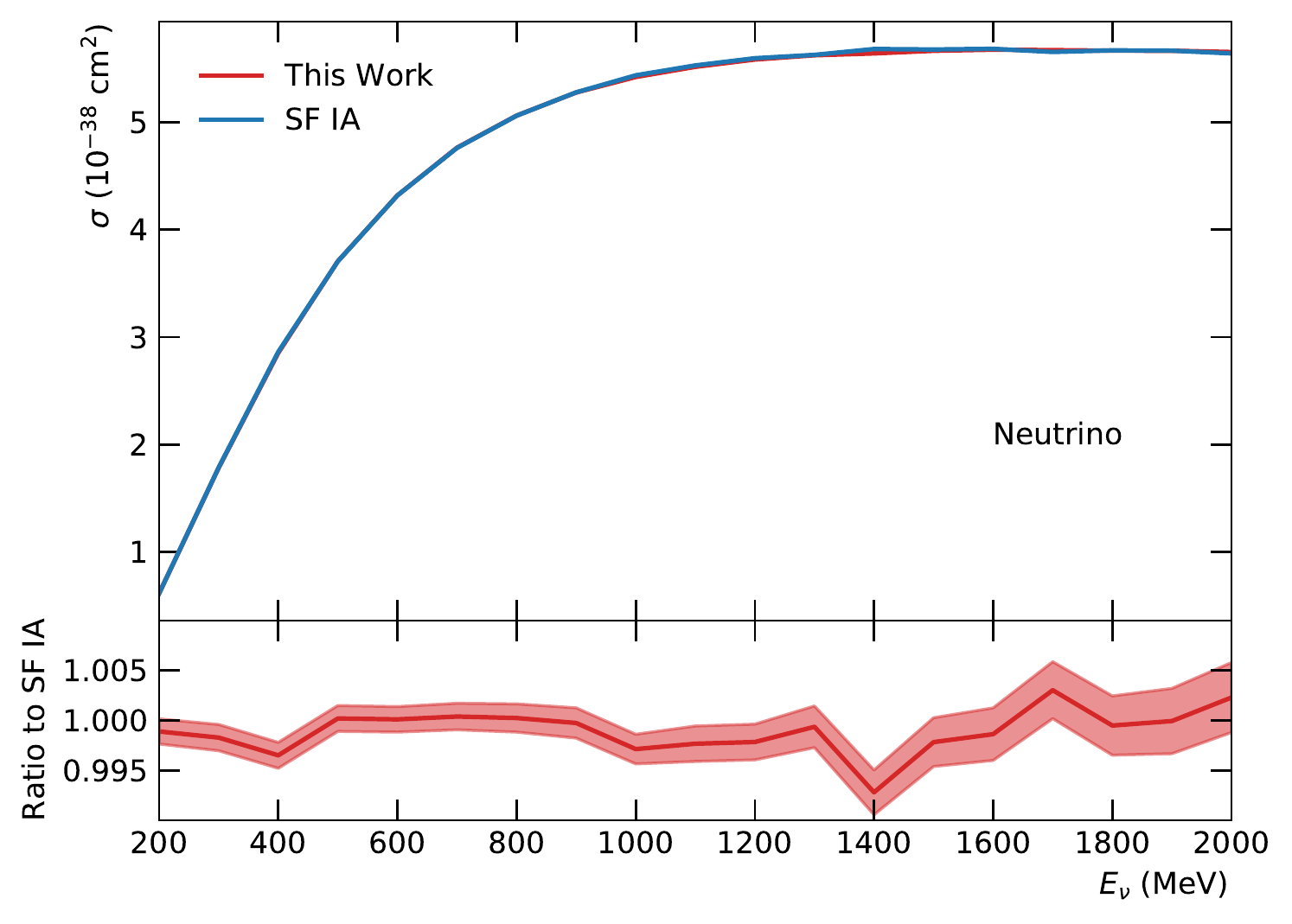}
    \hfill
    \includegraphics[width=0.45\textwidth]{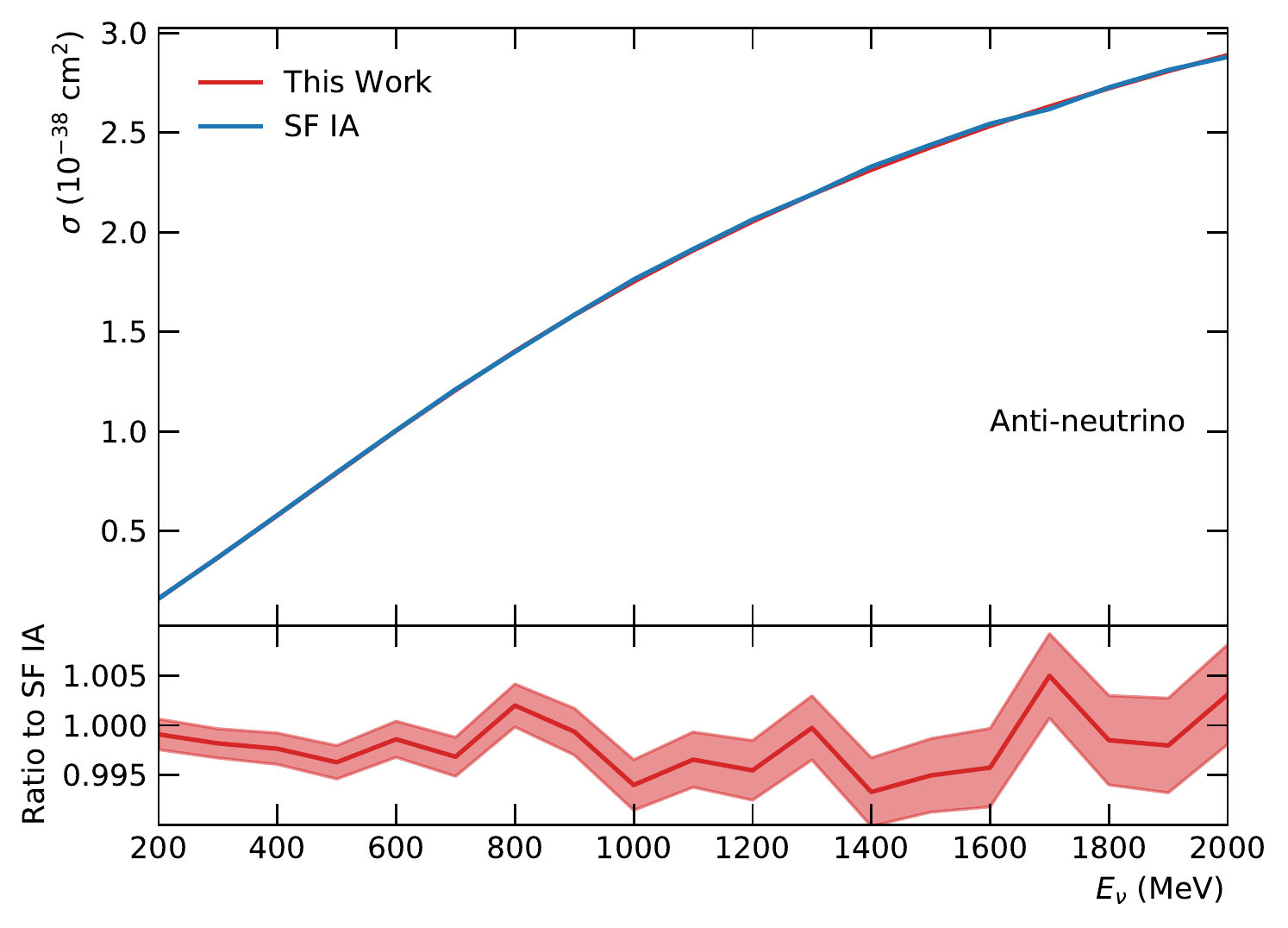}
    \caption{
    A comparison between our calculation and the SF IA based on Ref.~\cite{Rocco:2018mwt} for the total cross section for charged current
    scattering off a Carbon nucleus versus neutrino energy. On the left is the neutrino channel,
    while on the right is the anti-neutrino channel.}
    \label{fig:CC_total_xsec}
\end{figure}

As shown in Fig.~\ref{fig:CC_total_xsec}, the calculation using the current approach agrees (up to statistical uncertainties) with the calculation based on Ref.~\cite{Rocco:2018mwt}.

\begin{figure}
    \centering
    \includegraphics[width=0.45\textwidth]{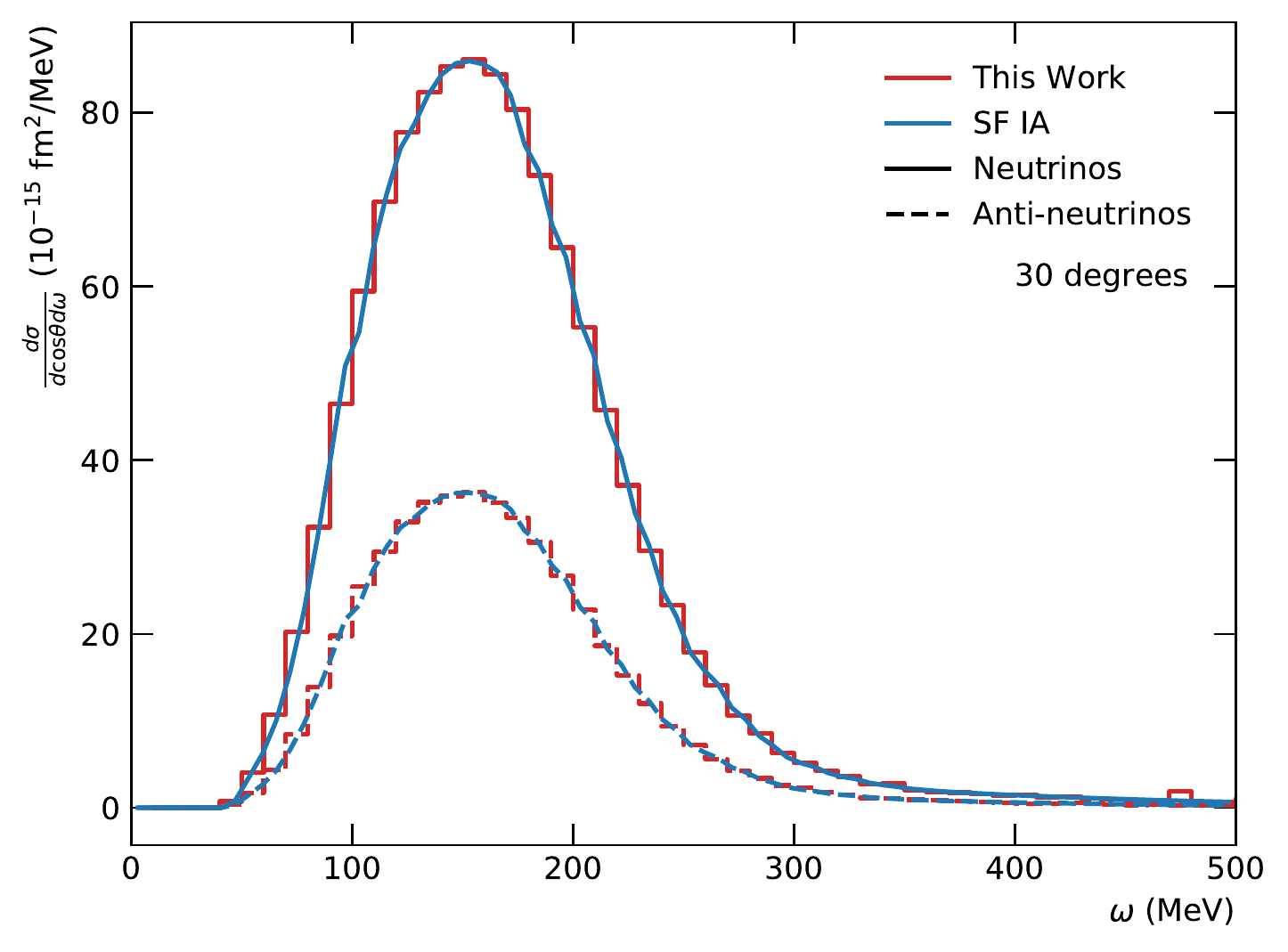}
    \hfill
    \includegraphics[width=0.45\textwidth]{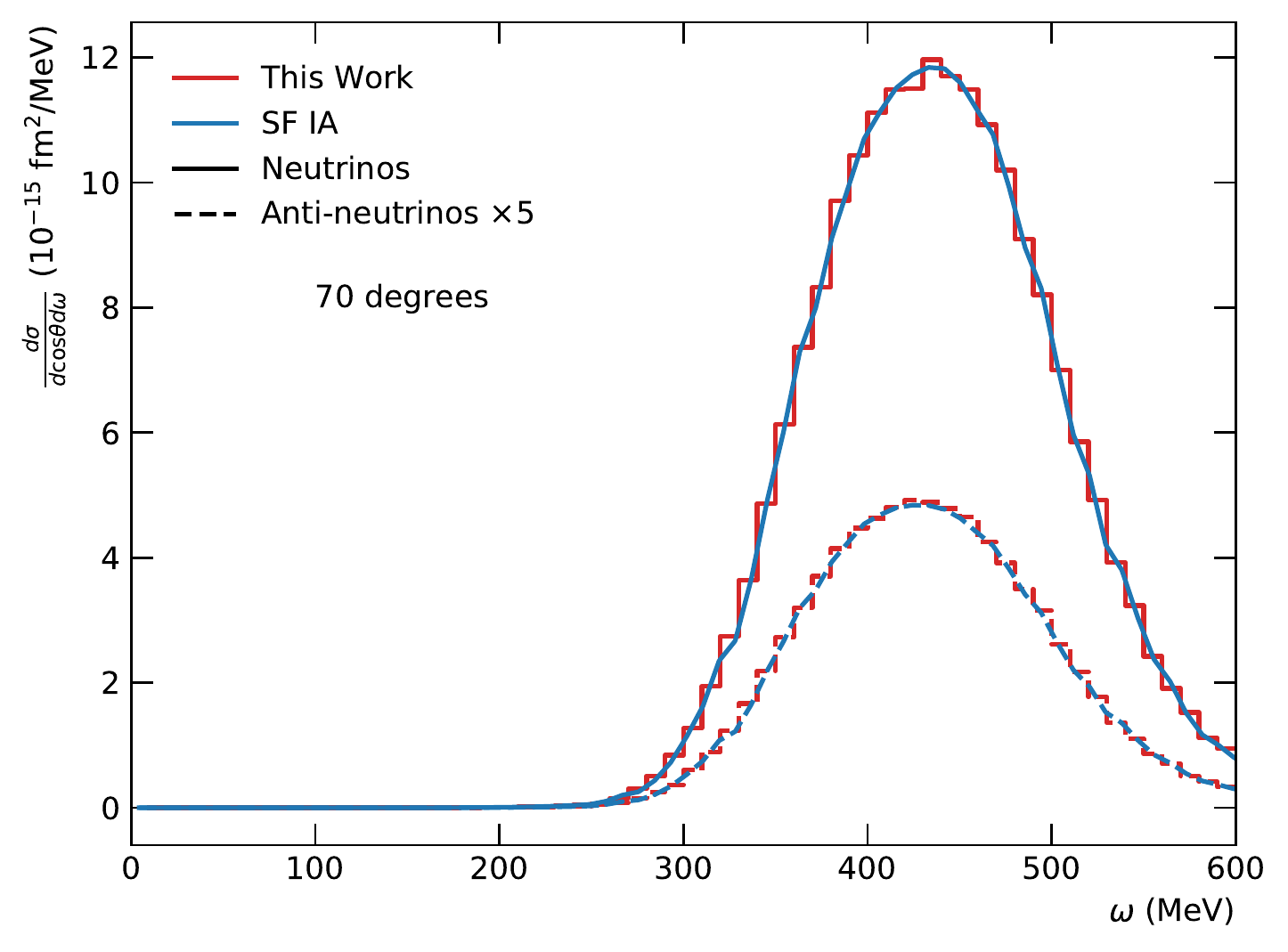}
    \caption{A comparison between our calculation and the SF IA based on Ref.~\cite{Rocco:2018mwt} for
    the double differential cross section for a 1 GeV neutrino on a Carbon target, with a fixed outgoing lepton angle of 30 degrees on the left and 70 degrees on the right.}
    \label{fig:CC_differential_nu}
\end{figure}

Additionally, the differential cross section is compared in Fig.~\ref{fig:CC_differential_nu} for the CC
interaction at a fixed outgoing lepton angle of $30^\circ$ (left) and $70^\circ$ (right) for an incoming 1 GeV neutrino and anti-neutrino. Here we again see consistency between our method and that of Ref.~\cite{Rocco:2018mwt}.

\subsection{Neutrino Tridents}\label{subsec:tridents}

\begin{figure}[ht!]
    \centering
    \begin{tikzpicture}
  \begin{feynman}
    \vertex (a) {\(\nu_\mu\)};
    \vertex [right=of a] (b);
    \vertex [right=of b] (f1) {\(\nu_{\mu}\)};
    \vertex [below=of b] (c);
    \vertex [right=of c] (f2) {\(e^{+}\)};
    \vertex [below=of c] (d);
    \vertex [right=of d] (f3) {\(e^{-}\)};
    \vertex [below=of d] (e);
    \vertex [left=of e] (i1) {\(N\)};
    \vertex [right=of e] (f4) {\(N\)};

    \diagram* {
      (a) -- [fermion] (b) -- [fermion] (f1),
      (b) -- [boson, edge label'=\(Z\)] (c),
      (c) -- [anti fermion] (f2),
      (c) -- [fermion] (d) -- [fermion] (f3),
      (d) -- [boson, edge label'=\(Z / \gamma \)] (e),
      (i1) -- [fermion] (e) -- [fermion] (f4),
    };
  \end{feynman}
\end{tikzpicture}
\hspace{2cm}
    \begin{tikzpicture}
  \begin{feynman}
    \vertex (a) {\(\nu_\mu\)};
    \vertex [right=of a] (b);
    \vertex [right=of b] (f1) {\(\nu_{\mu}\)};
    \vertex [below=of b] (c);
    \vertex [right=of c] (f2) {\(e^{-}\)};
    \vertex [below=of c] (d);
    \vertex [right=of d] (f3) {\(e^{+}\)};
    \vertex [below=of d] (e);
    \vertex [left=of e] (i1) {\(N\)};
    \vertex [right=of e] (f4) {\(N\)};

    \diagram* {
      (a) -- [fermion] (b) -- [fermion] (f1),
      (b) -- [boson, edge label'=\(Z\)] (c),
      (c) -- [fermion] (f2),
      (c) -- [anti fermion] (d) -- [anti fermion] (f3),
      (d) -- [boson, edge label'=\(Z / \gamma \)] (e),
      (i1) -- [fermion] (e) -- [fermion] (f4),
    };
  \end{feynman}
\end{tikzpicture}
    \caption{The two Feynman diagrams representing the neutrino trident cross section.}
    \label{fig:feynman_tridents}
\end{figure}
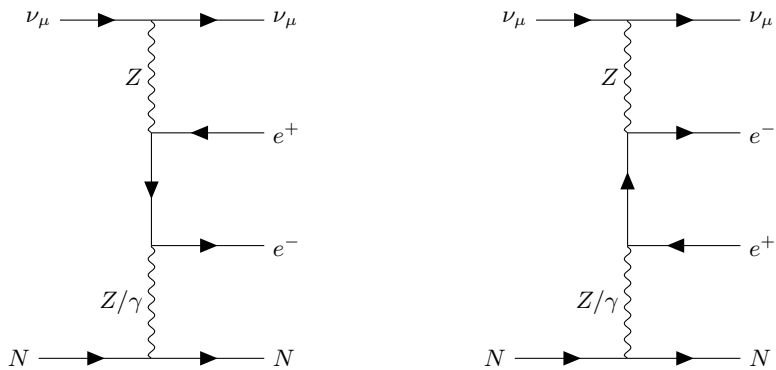

\begin{figure}
    \centering
    \includegraphics{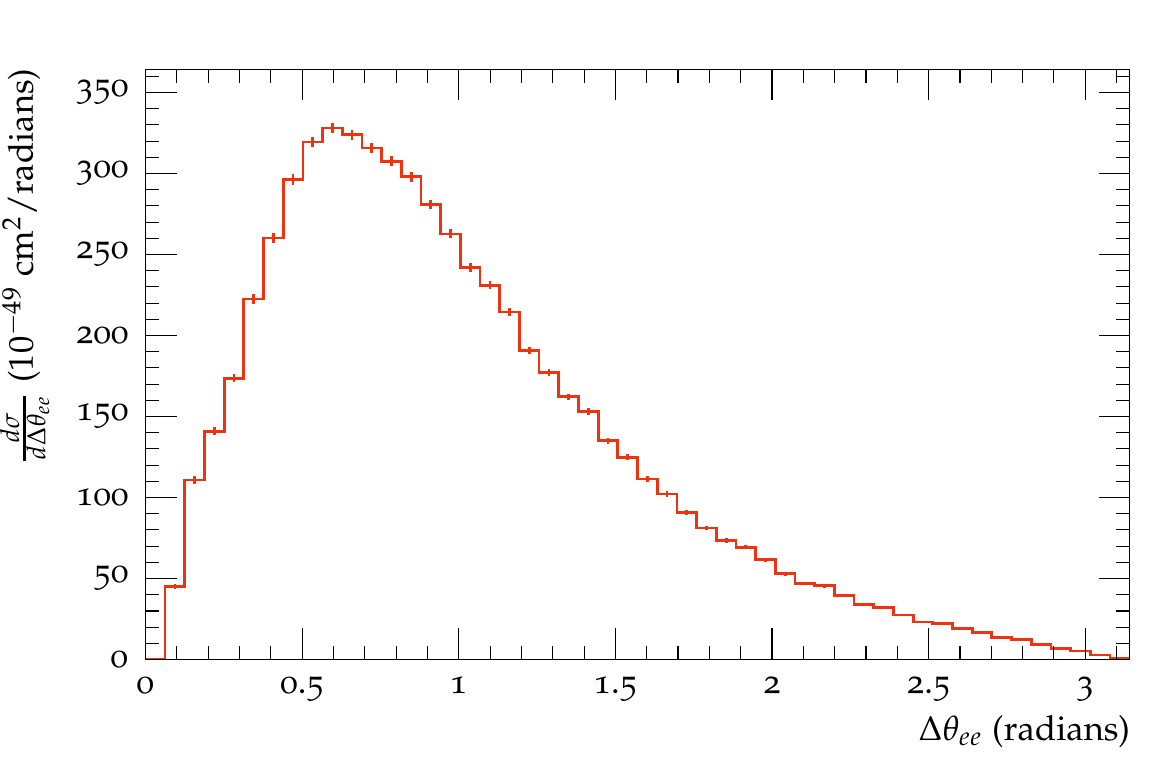}
    \caption{The opening angle between the two electrons in the neutrino trident process with an incoming neutrino beam of 1 GeV. A cut is placed on the minimum opening angle of $5^\circ$, and the electrons are required to have a minimum energy of 30 MeV.}
    \label{fig:trident_opening_angle}
\end{figure}

\begin{figure}
    \centering
    \includegraphics[width=0.45\textwidth]{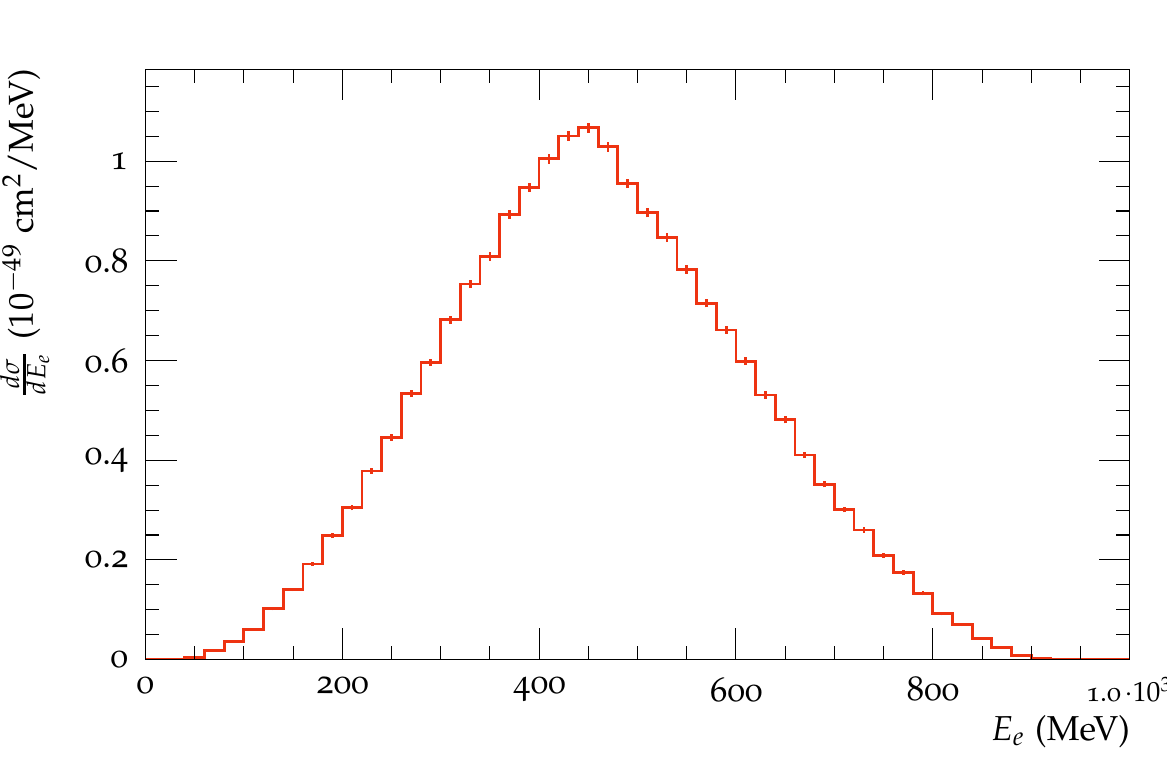}
    \hfill
    \includegraphics[width=0.45\textwidth]{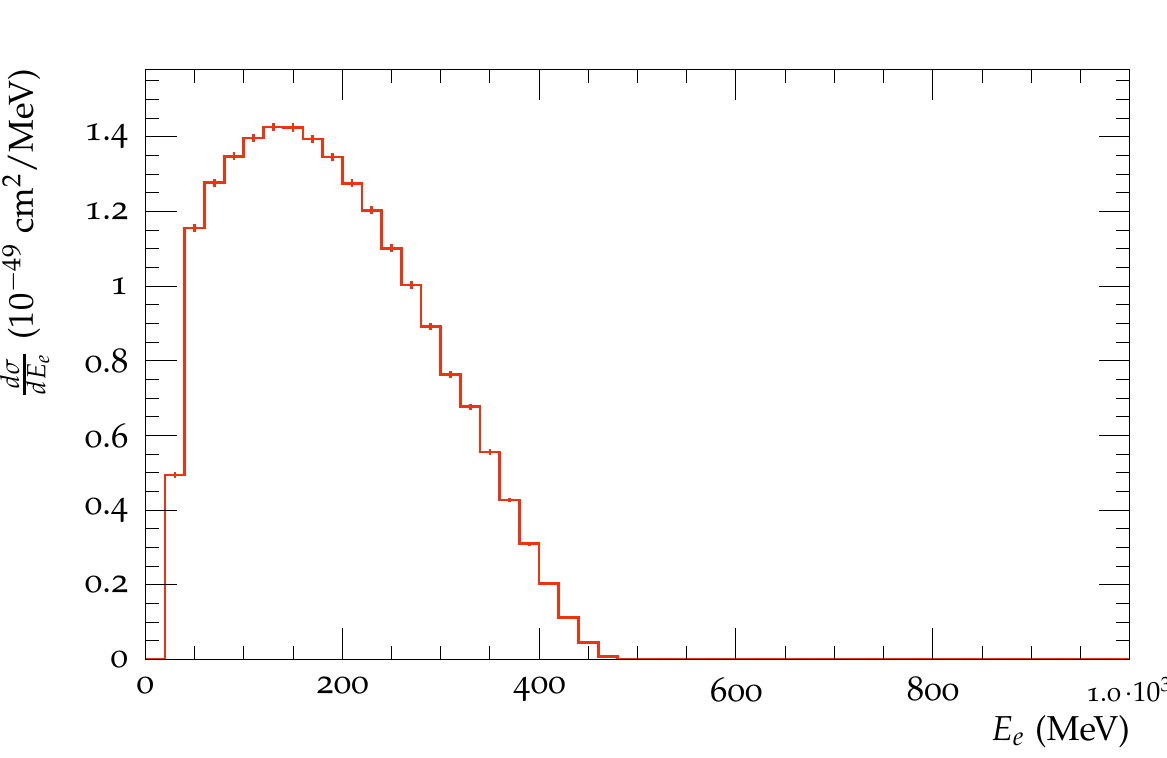}
    \caption{The leading (left) and sub-leading (right) energy for the outgoing electrons in the neutrino trident process with an incoming neutrino beam of 1 GeV. A cut is placed on the minimum opening angle between the electrons of $5^\circ$, and the electrons are required to have a minimum energy of 30 MeV.}
    \label{fig:trident_electron_energy}
\end{figure}

\begin{figure}
    \centering
    \includegraphics{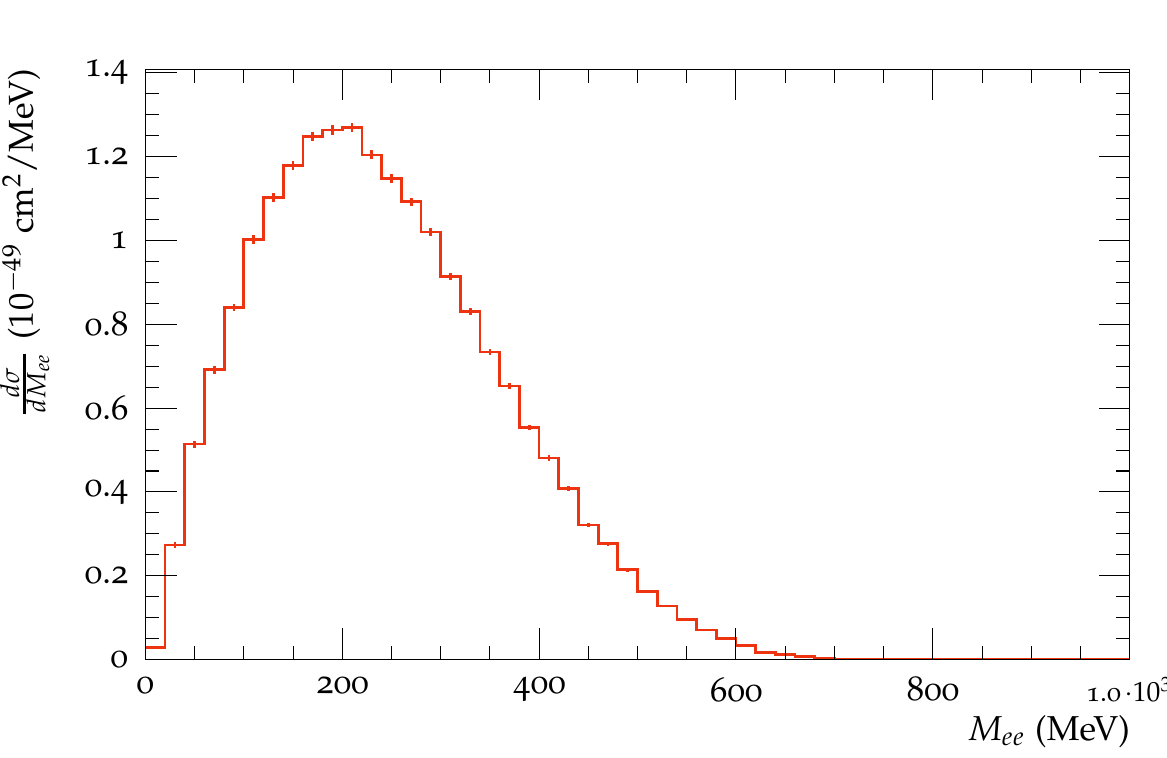}
    \caption{The invariant mass of the electron pair from the neutrino trident process with an incoming neutrino beam of 1 GeV. A cut is placed on the minimum opening angle between the electrons of $5^\circ$, and the electrons are required to have a minimum energy of 30 MeV.}
    \label{fig:trident_invariant_mass}
\end{figure}

To demonstrate that this generator is not limited to $2 \to 2$ process only and that it can handle interference terms, we consider the neutrino trident process $\nu_\mu\, ^{12}{\rm C} \to \nu_\mu e^{+} e^{-} X$ with a fixed neutrino beam energy of 1 GeV, including both the $Z$ and photon interactions with the nucleon. Additionally, this process is an important background to understand for multiple lepton final state BSM explanations of the MiniBooNE excess. The Feynman diagrams for this process can be
found in Fig.~\ref{fig:feynman_tridents}. In order to regulate the electron propagator pole, we set a minimum opening angle between the two electrons of $5^\circ$. In addition to this cut, we also require that the electrons have a minimum energy of 30 MeV, and have an angle from the neutrino beam axis greater than $10^\circ$. The electroweak parameters and form factors are identical to those used in the electron and neutrino scattering results section. With this setup, we obtain a total cross section of $3.973\times 10^{-14} \pm 2.764\times 10^{-17}$ nb, which is consistent with the results of Ref.~\cite{Ballett:2018uuc}. In addition, we show the results for the electron pair opening angle, the leading electron energy, the sub-leading electron energy, and the invariant mass of the electron pair. The electron pair opening angle can be seen in Fig.~\ref{fig:trident_opening_angle}, and is defined as the angle between the two electrons. This observable is important in understanding the ability of the next-generation experiments to observe this process for the first time based on their resolution for separating the two electrons from a single electron. The leading and sub-leading electron energy can be seen in Fig.~\ref{fig:trident_electron_energy}. In comparing the leading and sub-leading energies, we see that one electron tends to be significantly softer than the other electron. Finally, the electron pair invariant mass can be seen in Fig.~\ref{fig:trident_invariant_mass}. This is an important observable to distinguish neutrino trident processes from BSM scenarios that have a pair of electrons in the final state, such as Ref.~\cite{Bertuzzo:2018itn}. In the case of Ref.~\cite{Bertuzzo:2018itn}, the $Z'$ will create a bump in the invariant mass spectrum, which would be easily distinguishable from the standard neutrino trident prediction. We leave a detailed analysis of separating the standard neutrino trident from $Z'$ models to a future work.

\section{Conclusions}\label{sec:conculsions}

We developed a novel event generation framework for the automated simulation of neutrino scattering at next-generation neutrino experiments. The framework takes inspiration from similar tools developed for the automatic simulation of events for the LHC community. The major complication that does not exist at the LHC is the handling of the nuclear physics effects, which we address by interfacing to a dedicated code for nuclear physics models in the quasielastic scattering region.
Adding in the two-body currents, resonance production, shallow inelastic scattering, and deep inelastic scattering contributions is a logical extension of this work. This would require updating the handling of the initial states for the nuclear side, modifying the energy conserving delta function such that the phase space techniques described here can be applied, and implementing the additional nuclear effects. We leave this to a future work.

We demonstrate that we reproduce the expected results for the commonly studied processes of electron scattering and neutrino scattering off nuclei. 
%A constant offset between our results and the reference calculation can be attributed to the different ways of handling electroweak input parameters and the $W$ boson propagator in the two different frameworks. 
Furthermore, we demonstrate the ability of the framework to compute processes beyond $2 \to 2$ scattering by studying neutrino tridents. This process is important for multi-lepton final state explanations of the MiniBooNE excess. We show a variety of differential distributions demonstrating that this framework is capable of simulating full-differential events for subsequent analysis in the experimental simulation pipelines.

With the development of our new event generator, it becomes straightforward to study possible beyond Standard Model physics scenarios in a rigorous manner. Since the results obtained from the generator are fully differential in the many-body phase space and include the complete nuclear effects, our framework can assist the experiments in defining improved search strategies to separate various BSM scenarios from the Standard Model and from each other. We leave the details of this procedure to a future work.

The source code for our event generator will be provided upon request, and made public upon publication of this work.

\section{Acknowledgments}
We thank Pedro Machado, William Jay, Alessandro Lovato, and Gil Paz for useful discussions and advice throughout the development of this tool. This manuscript has been authored by Fermi Research Alliance, LLC under Contract No. DE-AC02-07CH11359 with the U.S. Department of Energy, Office of Science, Office of High Energy Physics.

\appendix
\section{Natively supported Lorentz structures}
\label{sec:lorentz_structures}
To exemplify the generality of our code, we list in this appendix the
nontrivial expressions for the natively implemented Lorentz structures,
not including those that correspond to simple contractions of external
polarization vectors or spinors.
The building blocks given here can be extended to nearly arbitrary 
interactions (also including higher-point functions) by means of 
the interface to FeynRules and UFO published in Ref.~\cite{Hoche:2014kca}. 
\newcommand{\dst}{\displaystyle}
\newcommand{\rbr}[1]{\left( #1\right)}
\newcommand{\smallfeynmf}{
  \fmfset{thick}{1.25thin}
  \fmfset{arrow_len}{2mm}\fmfset{curly_len}{1.5mm}
  \fmfset{wiggly_len}{1.5mm}\fmfset{decor_size}{2mm}
  \fmfset{dot_len}{1mm}\fmfset{dot_size}{2thick}
  \fmfset{dash_len}{1.5mm}
  \unitlength=.5mm}
\newcommand{\vxxs}[4]{\parbox{2cm}{\begin{center}
  \smallfeynmf\begin{fmfgraph*}(25,20)
    \fmftop{i,j}
    \fmfbottom{ij}
    \fmf{#1}{i,v1,j}
    \fmf{#2,tension=1.5}{v1,v2}
    \fmf{phantom,tension=5}{ij,v2}
    \fmfv{d.size=0,l.d=3.0,l.a=90,l=$\dst #3$}{i}
    \fmfv{d.size=0,l.d=3.0,l.a=90,l=$\dst #4$}{j}
    \fmfv{d.shape=circle,d.size=3,l.a=0}{v2}
  \end{fmfgraph*}\end{center}}}
\newcommand{\vxxl}[4]{\parbox{2cm}{\begin{center}
  \smallfeynmf\begin{fmfgraph*}(25,20)
    \fmftop{i,j}
    \fmfbottom{ij}
    \fmf{#1}{v1,i}
    \fmf{#2}{j,v1}
    \fmf{#1,tension=1.5}{v2,v1}
    \fmf{phantom,tension=5}{ij,v2}
    \fmfv{d.size=0,l.d=3.0,l.a=90,l=$\dst #3$}{i}
    \fmfv{d.size=0,l.d=3.0,l.a=90,l=$\dst #4$}{j}
    \fmfv{d.shape=circle,d.size=3,l.a=0}{v2}
  \end{fmfgraph*}\end{center}}}
\newcommand{\vxxr}[4]{\parbox{2cm}{\begin{center}
  \smallfeynmf\begin{fmfgraph*}(25,20)
    \fmftop{i,j}
    \fmfbottom{ij}
    \fmf{#2}{v1,i}
    \fmf{#1}{j,v1}
    \fmf{#1,tension=1.5}{v1,v2}
    \fmf{phantom,tension=5}{ij,v2}
    \fmfv{d.size=0,l.d=3.0,l.a=90,l=$\dst #3$}{i}
    \fmfv{d.size=0,l.d=3.0,l.a=90,l=$\dst #4$}{j}
    \fmfv{d.shape=circle,d.size=3,l.a=0}{v2}
  \end{fmfgraph*}\end{center}}}
\newcommand{\vxxxl}[6]{\parbox{2cm}{\begin{center}
  \smallfeynmf\begin{fmfgraph*}(25,20)
    \fmftop{i,j,k}
    \fmfbottom{ijk}
    \fmf{#1}{v1,i}
    \fmf{#2}{v1,j}
    \fmf{#3}{v1,k}
    \fmf{#1,tension=3}{v1,v2}
    \fmf{phantom,tension=5}{ijk,v2}
    \fmfv{d.size=0,l.d=3.0,l.a=90,l=$\dst #4$}{i}
    \fmfv{d.size=0,l.d=3.0,l.a=90,l=$\dst #5$}{j}
    \fmfv{d.size=0,l.d=3.0,l.a=90,l=$\dst #6$}{k}
    \fmfv{d.shape=circle,d.size=3,l.a=0}{v2}
  \end{fmfgraph*}\end{center}}}
\begin{align}\label{eq:cpl_ffvr_a}
    \vxxl{fermion}{photon}{\bar u}{j}
    =\bar{u}j^\mu\gamma_\mu \frac{1-\gamma^5}{2} 
    &=\rbr{0,0,\bar{u}_0j^--\bar{u}_1j_\perp,
    -\bar{u}_0j_\perp^*+\bar{u}_1j^+}\;,
\end{align}
\begin{align}\label{eq:cpl_ffvr_b}
  \vxxr{fermion}{photon}{j}{v}
  =j^\mu\gamma_\mu \frac{1-\gamma^5}{2} v
    &=\left(\begin{array}{c}0\\0\\
     j^+v_0+j_\perp^*v_1\\j_\perp v_0+j^-v_1\end{array}\right)\;,
\end{align}
\begin{align}\label{eq:cpl_ffvr_c}
  &\vxxs{fermion}{photon}{\bar u}{v}
  =\bar{u}\gamma^\mu \frac{1-\gamma^5}{2} v 
    =\left(\begin{array}{c}
      \bar{u}_0v_2+\bar{u}_1v_3\\
      \bar{u}_0v_3+\bar{u}_1v_2\\
      i\left(\bar{u}_1v_2-\bar{u}_0v_3\right)\\
      \bar{u}_0v_2-\bar{u}_1v_3\end{array}\right)\;,
\end{align}
\begin{align}\label{eq:cpl_ffvl_a}
  \vxxl{fermion}{photon}{\bar u}{j}
  =\bar{u}j^\mu\gamma_\mu \frac{1+\gamma^5}{2}
    &=\rbr{\bar{u}_2j^++\bar{u}_3j_\perp,
    \bar{u}_2j_\perp^*+\bar{u}_3j^-,0,0}\;,
\end{align}
\begin{align}\label{eq:cpl_ffvl_b}
  \vxxr{fermion}{photon}{j}{v}
  =j^\mu\gamma_\mu \frac{1+\gamma^5}{2} v
    &=\left(\begin{array}{c}
    j^-v_2-j_\perp^*v_3\\-j_\perp v_2+j^+v_3\\
    0\\0\end{array}\right)\;,
\end{align}
\begin{align}\label{eq:cpl_ffvl_c}
  &\vxxs{fermion}{photon}{\bar u}{v}
  =\bar{u}\gamma^\mu \frac{1+\gamma^5}{2} v
    =\left(\begin{array}{c}
    \bar{u}_2v_0+\bar{u}_3v_1\\
    -\bar{u}_2v_1-\bar{u}_3v_0\\
    i\left(\bar{u}_2v_1-\bar{u}_3v_0\right)\\
    -\bar{u}_2v_0+\bar{u}_3v_1\end{array}\right)\;,
\end{align}
\begin{align}
  \vxxs{photon}{photon}{j(p)}{j'(q)}
  =\Gamma^{\mu\lambda\kappa}\rbr{p,q}j_\lambda j'_\kappa
    &=jj'\rbr{p-q}^\mu
    +j'\rbr{2q+p}j^\mu
    -j\rbr{2p+q}j'^\mu\;,
\end{align}
\begin{align}
  \vxxxl{photon}{photon}{photon}{j}{j'}{j''}
  =(2g^{\mu\nu}g^{\lambda\kappa}-g^{\mu\lambda}g^{\nu\kappa}
    -g^{\mu\kappa}g^{\nu\lambda})j_\nu j'_\kappa j''_\lambda
    &=2(j'j'')j^\mu
    -(jj')j''^\mu
    -(jj'')j'^\mu\;,
\end{align}

\end{fmffile}

\bibliography{biblio}
\end{document}